**The Galactic origin for the borders in the Earth history.**
R.S. Nigmatzyanov, mingeoasinkem@rambler.ru.

The Galactic origin for the borders in the Earth history.
The external galactic key factor for developing of massive geochronological boundaries, as well as mass extinctions (ME) is proved based on chronological relationships of the impulses of globally short processes. External galactic key factor is also proved based on the coincidence of the scales of their developing processes, their regular periodicity in the formation of the boundaries of the Phanerozoic epochs and the presence of a unifying complex of cause-effect relationships for boundary processes. The hypotheses of the impact origin of ocean basins and plumes are substantiated, the assumptions about the unsteady nature of the Spiral Arms and the rotational acceleration of the Milky Way's core are confirmed.
*Key words:* impact origin of ocean basins and plumes; the rotational acceleration of the core and widening Galaxy Spiral Arms.

**Introduction.**
The collected data from numerous studies indicate that the globally long gradual processes are periodically interrupted by chronologically related massive transformations at the borders of geological history for a relatively short time within 1-6 million years [*1; 2; 19*]. These changes have impulsive nature and appear spasmodically in the face of previous smooth changes as the following:
   - huge astronomical bodies' impact,
   - neogenesis of ocean (deep-sea) basins and plumes,
   - falling of sea level and decreasing oxygen,
   - sharp climate changes,
   - mass extinction (ME),
   - changes of the equator's pale and Earth rotational rate,
   - tectogenesis flashes and changes in inversions characters,
   - planet cooling spikes,
   - impulse magmatism of plume and basalts.

Geochronological partition of the Earth's history into stages is carried out in accordance with fundamental environmental changes (mainly on the subsequent long-term restoration-distribution of biota [*1*]). For example, an iridium anomaly is officially approved as an edge marker for the Mesozoic and Cenozoic eras that presumably fixes the moment of a huge asteroid's impact [*1*] (or a relatively smooth release of accumulated cosmic dust by cover glaciers). In total there are up to 60 different globally labeled events distinguished in Phanerozoic according to [*1; 6; 12; 24*] (*Fig.1*).

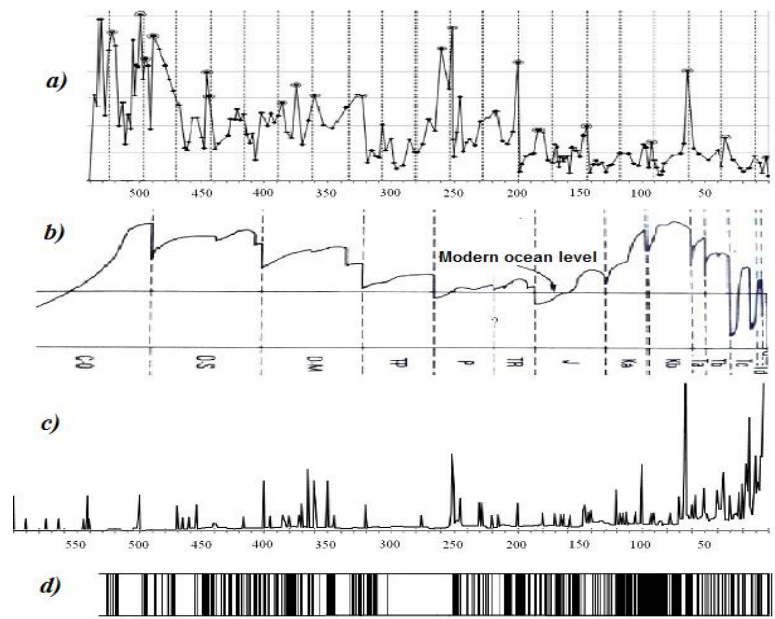

*Fig.1. a) Phanerozoic extinction rate [17], b) eustatic curve for ocean level changes [24], c) intensity of impact events (by [18]), d) Phanerozoic magnetostratigraphic scale (E.F. Molostofsky, 2006).*

The most frequently cited in scientific literature frequency intervals of events' occurrence for differing scales are 10-80 and 110-290 million years [*1; 6; 10; 12; 13; 19; 24; 25*], which approximately corresponds to Stille (*H.W. Stille*) cycles 30-45 million years and Bertrand cycles (*M.A. Bertrand*) of 150-200 million years [*3*]. Those numbers correspond to durations of epochs, periods of up to 100 million years and an era of geochronological scale of 155-375 million years [*8; 9*]; the turnover of the Galactic Center objects - 24 million years and the Galactic year - 160-275 million years [*2*] as well. A discrepancy for the periodicities of various processes among different researchers can be caused by inconsistency over a wide range of the boundaries dating for geological time scales by [*1; 2; 8; 9*] (*Chart.1; Fig.2*).

*Chart 1. An example of a scale in the range of possible values of the boundaries of epochs (by [8; 9]) with a decrease in the duration of each subsequent era by 0.7 million years.*

| boundaries of epochs for geological time scales | minimum value, million years ago | maximum value, million years ago | estimated border, million years ago | duration of eras, million years |
|---|---|---|---|---|
| Q / N | 2,6 | 1,0 | 1,0 | Q = 11,5 |
| Pli / Mio | 13,0 | 5,0 | 13,2 | Pli = 12,2 |
| Mio / Oli | 27,0 | 22,5 | 26,1 | Mio = 12,9 |
| Oli / Eoc | 40,0 | 33,8 | 39,7 | Oli = 13,6 |
| Eoc / Pal | 60,0 | 53,0 | 54,0 | Eoc = 14,3 |
| Pal / Sen | 70,0 | 63,0 | 69,0 | Pal = 15,0 |
| Sen / | 90,3 | 80,0 | 84,7 | Sen = 15,7 |
| $K_2 / K_1$ | 110,0 | 95,0 | 101,1 | $(K_2 - Sen) = 16,4$ |
| / Neo | 131,8 | 118,0 | 118,2 | $(K_1 - Neo) = 17,1$ |
| $K_1 / J_3$ | 149,5 | 125,0 | 136,0 | Neo = 17,8 |
| $J_3 / J_2$ | 165,2 | 149,0 | 154,5 | $J_3 = 18,5$ |
| $J_2 / J_1$ | 188,0 | 172,0 | 173,7 | $J_2 = 19,2$ |
| $J_1 / T_3$ | 213,0 | 180,0 | 193,6 | $J_1 = 19,9$ |
| $T_3 / T_2+T_1$ | 235,0 | 200,0 | 214,2 | $T_3 = 20,6$ |
| $T_2+T_1 / P_2$ | 254,0 | 225,0 | 235,5 | $T_2+T_1 = 21,3$ |
| $P_2 / P_1$ | 272,8 | 240,0 | 257,5 | $P_2 = 22,0$ |
| $P_1 / C_3$ | 300,0 | 270,0 | 280,2 | $P_1 = 22,7$ |
| $C_3 / C_2$ | 308,2 | 300,0 | 303,6 | $C_3 = 23,4$ |
| $C_2 / C_1$ | 328,0 | 316,8 | 327,7 | $C_2 = 24,1$ |
| $C_1 / D_3$ | 368,0 | 345,0 | 352,5 | $C_1 = 24,8$ |
| $D_2 / D_1$ | 400,2 | 370,0 | 378,0 | $D_3+D_2 = 25,5$ |
| $D_1 / S$ | 422,4 | 395,0 | 404,2 | $D_1 = 26,2$ |
| $S / O_2$ | 446,0 | 425,0 | 431,1 | $S = 26,9$ |
| $O_2 / O_1$ | 473,4 | 445,0 | 458,7 | $O_2 = 27,6$ |
| $O_1 / Є_3$ | 510,0 | 487,0 | 487,0 | $O_1 = 28,3$ |
| $Є_3 / Є_2$ | 530,0 | 497,0 | 516,0 | $Є_3 = 29,0$ |
| $Є_2 / Є_1$ | 549,0 | 509,0 | 545,7 | $Є_2 = 29,7$ |
| $Є_1 / V$ | 600,0 | 534,0 | 576,1 | $Є_1 = 30,4$ |

*In this example (Chart.1; Pic.2) the border $T_1/T_2$ is excluded as a possible marker of second impulse of events at the border of P/T by [1] within 1-6 million years (by [1; 2; 19]), senon and neocom are considered as sub-epochs of Chalk period; Perm, Devon and Ordovician have a two-term division; Silurium is not divided (by [8]).*

The scale of events has huge interconnections between each other. For example, the boundary between the largest Paleozoic and Mesozoic eras Pz/Mz (P/T eras) of the Phanerozoic correspond to the following features:
- huge two-pulse mass extinction of up to 95% for all species (by [*1; 11; 8*]);
- completion of a long-term global glaciation (up to 80-90 million years) which caused a geocratic climate on the continents [*11*] and the absence of significant glaciations until the end of the Paleogene [*13*];
- the largest fall in sea level with a break in the Tethys sedimentary sections, closing-isolation of the Late Precambrian-Early Paleozoic oceans and strip shaped Pangea supercontinent (up to 7 thousand km wide) from pole to pole (by [*1; 10; 12; 13; 25*]);

- oxygen drop to the Early Paleozoic values, followed by a rapid increase to a historical maximum of 200-125 million years ago [*11; 21; 25*];

- formation of the Panthalassa superocean by [*10*], including the antipodal Pacific and Pre-Atlantic marginal super cavities, as well as a significant change in the configuration of the Pra-Tethys [*10; 11*];

- two impulses for the numerous formation of diatreme ("blow hole"), with the supposed thickness of explosive deposits is up to 1000 m [*15*], possibly the largest impact in the form of Panthalassa itself;

- a surge in planet's global cooling, which is expressed in the maximum impulses of magmatism in the Paleozoic followed by the formation of the largest Tunguska trappy province and Emeishan traps [*12; 15*];

- the hypothetical origin of antipodal superplumes beneath the Pacific Ocean and Africa, which are widely and chronologically associated with the Pacific and Pre-Atlantic basins of Panthalassa [*10; 15; 20*];

- fast and massive rotation of the planet's equator plane, possibly up to $90°$ [*21*];

- 90-degree reconstruction of the planetary tectonic plan with the formation of the folded structures of the Urals and the Appalachian Mountains from the crust of ancient abyssal depth; the beginning of the relative maximum for the absolute speeds of the continents in the range of 250-220 million years ago [*6; 10; 12; 13*];

- change of a superchron of reverse polarity - a relatively quiet geomagnetic field with extremely rare inversions in the range of 300-250 million years ago with doubled peaks of increase in the frequency of inversions between 260-210 million years ago [*6; 8*].

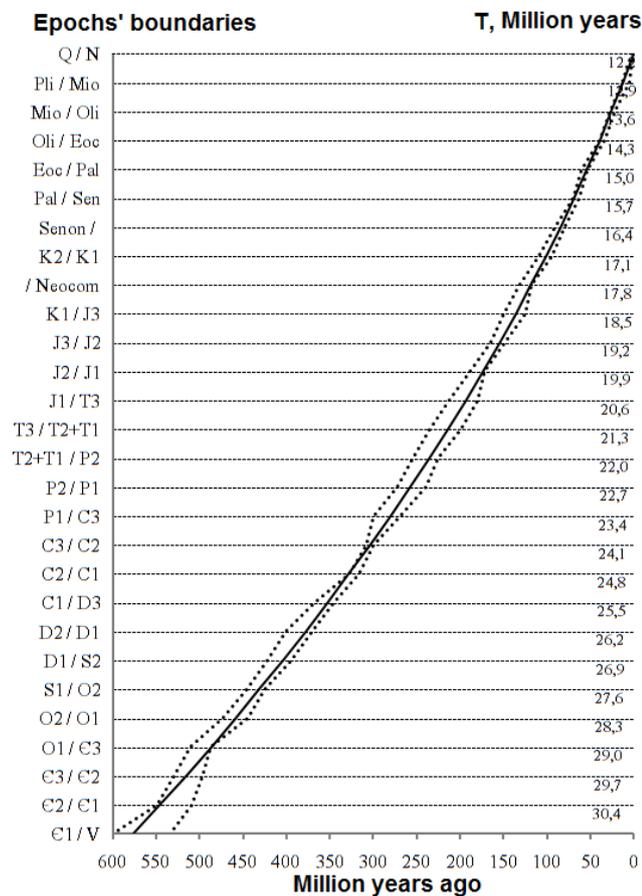

*Pic.2. The example of a possible arithmetic regression (solid midline) - periodicity from 30.4 to 12.2 million years in the range of maximum and minimum values of the Phanerozoic boundaries for different scales of geological time (by [8; 9]).*

***Conclusion:*** Comparing reoccurrence for various short-term processes in the Phanerozoic history, chronological relationships and regular periodicities, the coincidence of the scale and number of pulses of various boundary events, as well as the correlation between the amplitudes of boundary events and the scale of subsequent gradual long-term processes, we can indicate the cause-and-effect linkage between them. But these coincidences cannot justify the unified nature of the above-mentioned processes till establishing a unifying system of self-consistent relationships [*1; 10; 13*]. This work is aimed to establish such a system.

**Formulation of the problem.**

During comparing chronologically related short-term global processes it is assumed by many researchers, that mass extinctions are one of the consequences in the complex of catastrophic events at the turn of the epochs. In accordance with the law of diversity growth of the biosphere during a stable state of the environment, the taxonomic diversity of the biosphere must continuously increase (almost unlimited) without achieving stasis due to the evolutionary fragmentation of ecological niches [*1*]. Therefore, the biosphere is not considered as a possible cause of happened MEs or other globally short events. At present, a possibility of biota influence on the Phanerozoic climate as well as its influence on hydrocarbon deposits' formation and even on the oxygen level in the atmosphere is still an open question.

Impulses of large impacts could be the root cause of the cataclysms [*1; 11*], they cannot be the result of other planetary processes which are chronologically associated with them. But the main difficulty in considering sorts of relationships is a natural desire to explain everything by a single reason, ignoring the need for a systematic approach [*1*]. Therefore, in order to establish hierarchical relationships between the fast transformations of the planet that occurred in the past, the data analysis is carried out by the method of elimination from reverse: - from biota crises to the source. Dividing the boundary processes into immediate, triggering and primary ones (by [*1*]) and others which are not directly related to ME.

The main immediate abiotic causes of past boundary MEs of the external and internal hypothesis groups are thought to be the following boundary changes: 1) climate, 2) the composition of the atmosphere and hydrosphere, 3) the intensities of radiation as well as cosmic radiation [*1*]. The following factors (triggers for mass extinctions) are proposed as possible causes for the mentioned global changes at the turn of the epochs, chronologically associated with biota crises:

*1) causes of climate change:* - Decreased insolation due to atmospheric dustiness with external cosmic reasons or increased explosive volcanism. This assumption is not considered in this work, since there is data about the coincidence of the growth of biosphere diversity and on intensity increase of explosive volcanism. As well we do have data about mismatching of extinction scales in the Northern and Southern hemispheres with global dustiness of the atmosphere at the Cretaceous/Paleogene boundary (*K/P or Mz/Kz*) [*1; 3; 10; 19*]. This milestone event is chronologically, spatially and massively associated with the formation of deep-sea basins of the Arctic Ocean by [*10; 11*] (*Arctic-Eurasian*) ring *structure with a diameter up to 4800 km*) and the occurrence of the Icelandic (Arctic) plume. The extent of the Cretaceous/Paleogene (K/P or Mz/Kz) extinction in the Northern hemisphere is higher than in the Southern one [*10*], there is also data that the content of Ir of presumably asteroid origin decreases in the Southern direction from Denmark.

- Reducing the heat balance of the planet due to solar activity fluctuations [*1; 3*]. Decreasing potential (internal) energy of the planet is irreversible, and after glaciation the Earth could not return to its initial state only by solar radiation [*21*] due to increasing the surface reflectivity, thus the hypothesis of a fundamental decrease in the planet's temperature should include a subsequent recovery mechanism. Even doubling of estimated amount for kinetic energy of cosmic bodies that transforms into heat after impact with the Earth is not able to provide enough heat to recover, with the possible exception of deep penetration cases by [*16*]. Therefore, the hypothesis of climate cooling is considered as a short-term temperature decrease which took place directly on the surfaces of continents and continental basins with a decrease in atmospheric pressure.

- An atmospheric pressure decrease on the continents' surface after sharp drop for the boundary of hydrosphere and atmosphere in the first kilometers (by [*12; 21*]) and lowering of the bottom of the chionosphere (snow layer - glaciation layer) to the shelf. Such a large regression could be the reason for a global Antarctic-type glaciation, isolation of deep-sea basins with a decrease in the level of carbonate compensation in the first kilometers to the bottom of the basins (by [*10*]). This assumption includes hypotheses of climate cooling. The oceans are the reason for heating the atmosphere, as well as hypotheses of disturbance of water circulation. A globally rainless (xerothermal) environment, accompanied by soil degradation and depletion of food resources [*1*], could be resulted by even more massive regressions of the chionosphere roof below the surface of the continents: - Weil and others believe that the amplitudes of eustatic fluctuations are estimated only approximately [*10*]. This opinion is confirmed by the relatively low calculated amplitude of the Perm/Triassic regression in the work [*24*] (*Fig.1*).

*2) causes that made atmospheric and hydrospheric composition changed:* - Paroxysmal volcanism and release of huge quantities of carbon dioxide, nitrogen oxides and sulfur which presumably led to dramatic climate changes and impact on biota [*1; 10*]. The hypothesis is not considering the chronological coincidences of the growth of biosphere diversity and intensification of volcanism (by [*1; 3; 10; 19;*).

- Short-term nonoxygenated environments associated with eustatic fluctuations in sea level. The assumption of a toxic high oxygen content in the atmosphere as a possible cause of ME is excluded on the basis of the duration of growth periods and high oxygen level (by [*1; 25*]).

- The possibility of an isotopy pulse resulted by the growth of Mid-Ocean Ridges (MOR) and central uplifts is also excluded due to the length of the MOR growth to the stagnation of spreading and magmatism of existing uplifts and to plumes at the time of crises (by [*3; 19*]). The hypothesis of a short-term isotopy of the composition of ocean waters [*10*] during plume neoplasm can only be considered in combination with other processes resulted mainly from its effect on marine biota.

- Short-term changes in the atmospheric and ocean's chemistry after falling of space bodies, for example partially hydrocarboned comets [*2; 11; 16*].

- A sharp increase for toxic content of methane in the atmosphere after pressure loss followed by gas hydrates' release. As well as a sharp oxygen drop to globally anoxic conditions on the continents and epicontinental basins due to massive regressions of the hydrosphere and atmosphere (by [*10; 13*]).

*3) causes of changes in radiation intensities:* - The hypothesis of the cosmic radiation growth after decrease in intensity of the planet's magnetic field resulted by paleomagnetic inversions is also excluded: there is no data about fundamental changes in the Earth's geomagnetic field during paleomagnetic pole movements over the past 2.5 billion years.

- The assumption of rising radiation pressure after filling ocean rift zones is not considered due to temporary mismatches of these processes with ME [*1*].

- The increase of cosmic radiation intensity [*10*]: - due to ozone layer degradation after intensification of periodic pulses of hydrogen degassing of the mantle and core [*1*] during new plumes' rising;

- Due to regressions of the troposphere and ozone layer over several kilometers;

- After intensifications of external, to the solar system, influences on its orbit [*1*], for example, at periodic intersections with Spiral Arms [*2*] and nearby supernovas' explosions [*10*].

*Conclusion:* Since the above mentioned chronologically related direct causes of ME have various sources, this work consider interconnections within the trigger complex of causes from massive global regressions, occurrence of new depressions and impact events and does not exclude other external cosmic influences.

**Discussion.**

*Global regressions:* - glaciation is not considered as a cause of mass regressions due to small amplitudes of glacioeustatic oscillations at 60-200 m [*10; 13*];

- the World ocean's redistribution after possible rapid turns with regression in the equatorial regions in the history of planet's equatorial plane, resulted by the difference in the polar and equatorial radii for about 22 km, that could lead to transgression in the polar regions based on materials [*10; 12; 13*];

- the transgressions and smooth fluctuations of the sea level within an amplitude of 300-500 m could have been caused by gradual changes in the MOR system as well as the rise of the central uplifts and the ocean floor, but the above mentioned processes are not able to explain the sharp sea level falls by hundreds of meters (by [*24*]).

After excluding the above mentioned hypotheses, the only possible reason of mass instantaneous regressions could be the rapid emergence of new circular deep-sea depressions (by [*7*]) with spasmodic fundamental changes-rearrangements of the configurations and sizes of previously existing basins (by [*10; 12; 25*]). Upon condition that initially the neogenesis were much deeper than the basins of modern seas and oceans (located on the surfaces of magmatic holes with depths up to 2900 km):

- circulation stop and isolation stop for deep-water basins are possible if the hydrosphere decreases to the ocean's floor by 4-5 km after a parabolic depression, with a capacity close to the modern volume of ocean waters of about 1.4 billion km$^3$ - in the size of the Coral Sea with an area of >4 million km$^2$ or the giant Arabian-Nubian Ring Structure of 2.2 thousand km and up to a third of the diameters of the structures depths;

- global mountain glaciation of the continents is possible when the lower border of the chionosphere falls from a level of about 5 km in the equatorial belt [*21*] below the shelf, which is acceptable for depressions with a capacity of >3.8 billion km$^3$ - the size of the Sargasso Sea or giant Amazonian Ring Structure with a diameter about 3.2 thousand km at depths of 1100-950 km;

- forming sparse, globally dry oxygen-free envinronment, sharply continental arid climate of red sedimentation and evaporate sediment genesis of the Martian type on the continents and shelf is possible after the following: decreasing limits of non-solid water at -41°C existing in the lower temperature, decreasing the chionosphere roof by >10.5 km in the equatorial zone. After forming a depression sized of the North American Ring Structure with a diameter of 3.8 thousand km and deepen up to 1/3 of the structure's size. This assumption is indirectly confirmed by the existence of arid-glacial thyloids (by [3]), as well as an evidently paradoxical scaling mismatch of climate changes in the North and the South of the Late Paleozoic [25], possibly caused by the difference in the first kilometers of the chionosphere borders in the Northern and Southern hemispheres, as well as by fast "evaporation"-freezing of epicontinental pools.

- the ozone layer's drop below the shelf and the oceans floor with an increase in hard cosmic radiation is possible after regression of the atmosphere by >10-50 km, which might have been due to basins formation ranging from the North American Ring Structure to the Pacific basin. Available data about the Pacific Ocean allow us to consider its structure as isostatically (magmatically) filling ring depression: some researchers compares the global system of subduction zones in the early stages of the Pacific basin formation as a giant funnel about 18 thousand kilometers diameter and a depth below the surface of the outer core.

*Neogenesis of depressions.* Currently, probably the following mechanisms for the formation of deep-sea depressions are indicated: riftogenesis and spreading [10], rapid basification-oceanization-sinking of the continental-type crust after the rise of a fluid-containing mantle plume to the lithosphere, after destruction of existing lithosphere and significant changes in the mantle's composition and structure, identical to impact (by [7; 16]). Depressions are divided into two fundamentally different types: as long-term developing (due to deeper ring-shaped) intercontinental split basins with divergent passive aseismic margins of the Atlantic type, and as rapidly forming, shaped with relatively isometric ring basins with central uplifts - asthenodiapyres and subduction convergent active seismic boundaries of Pacific type [10; 15; 21].

Reasonable objections to the hypothesis of basification are caused by formation of plumes' uplifts but not depressions, as well as by the unresolved issue of the plumes' origin. Impact events [7] are proposed not to contradict other short-term processes with the hypothesis of deep-sea annular basins' origin, which can be confirmed by synchronous bombing, regressions (*Fig.1*) and deep-sea neoplasms at the turn of the epoch: - With maxima areas of epicontinental seas (i.e., with the completion of transgressions - moments of regressions) and the growth's pick of uplifts coincides the intervals of pulses of kimberlite formation [15] of presumably shock origin.

A significant part of the planet's deep-sea basins is observed to be as closed lowerings in the oceans' bottom with more or less isometric shape. They represent flat surfaces with deviations of hundredths or thousandths of a degree – what is called as abyssal plains, that is similar to isostatically aligned surfaces of impact craters on other planets and their satellites. They are different from the Earth only by the absence of a hydrosphere [7; 15]. Similar shock ring structures which were filled with mantle masses are also found on the Earth - these are Sudbury and Vredefort craters [16].

Collisions with Mars sized bodies (6.8 thousand km) which took place in the Earth could have led to the formation of round craters from 1 to 40 projectile's diameter (by [16; 22]) and up to 80-90% of the target itself (in accordance with the relative sizes of the largest crater on the Westa asteroid). But less viscous bodies are possibly had the same impact funnels as its diameter and greater initial depths than the radius of the body itself.

During excavation stage, the parts of future crater are being ejected at a speed exceeding the second cosmic one leading to erosion of the planet – which means a decrease in its mass [7; 16] and increasing bombardment impulses after fast fragments' falling. These velocities of large fragments are sufficient for breaking through the thin crust above the central uplifts of deep-sea basins as well as for penetration into weakly viscous diapirs accompanied by diatreme's channels' formation according to Barringer's hypothesis of deep penetration of space bodies (*D.M. Barringer, 1905*). According to reports, kimberlite magmas entered into formed deep tubular depressions of various sizes as it appeared even the deepest xenoliths of kimberlite diatremes turned out to be only fragments of the ancient oceanic crust.

After excavation, the depth of momentary transitional crater (that has not collapsed) quantifies to one third of the depression diameter [16]. On the longer term after collapse, an isostatic (*magmatic - auth.*) rise can be observed, due to which the crater is flattened to almost flat albedo structure [16]. The necessity of magmatic processes in a large shock "explosive" funnel as after the mantle's pressure drop, even without visible free water or impactogenic outlet channels' formation for subcrustal magma, so the rise of the bottom of the transitional crater are confirmed by real objects' observations [7; 16] and do not require justifying the "convective-subduction immersion" of decompressed material in a denser mantle to the boundary of the outer core. Comparisons of such gravitationally and magmatically aligned impact structures (sized about and over 70-100 km) can lead to invalid conclusions about a decrease in the depths of craters with an increase in their size according to materials [*7; 16*], and a decrease in projectile's penetration increased in impacts' speed and scale. As a way of example, in Moon craters (with a diameter of more than 1800 km) the molten rocks' volume exceeds the excavation zone [*16*].

When the mantle melt rises, it decompresses to 3.3-2.9 g/cm$^3$ and increases proportionally in its volume. Increasing in asthenolite volume can form central uplifts: all types of depressed sedimentary structures' surface and deep relief are shaped with asthenodiapirs of molten magma, which seems to determine their formation. Diapir uplifts are formed from center to edges that is similar to large ring-type depressions of the centrifugal tectonics — the "widening" of newly formed crust from rising central part along to the sides of basins [*10; 21*] and the subsequent burial in the form of the observed isometric "cold" high-speed pseudosubduction rings.

***Bombing impulses***. According to published data, the possible number of potential impactors in the Solar system as comets and asteroids within 8.0 g/cm$^3$ densities and a total mass more than $10^4$-$10^6$ solar masses exceeds $10^{12}$-$10^{15}$. But the erosion of the terrestrial planets compared with the giant planets indicates the asteroid belt as the main source of impacts, with a total mass comparable to the mass of Mars body sizes up to 1000 km and affordable collision speeds with the Earth is about 0.1-72.8 km/s.

Epochs of relatively small-impact events lasted up to the first tens of millions of years (*Figs.1 and 2*), with a huge number of potentially dangerous large objects in the immediate vicinity to the ecliptic plane and frequent impacts in the asteroid belt [*16*], that prove the out Solar system's nature of the considered large-scale pulses of destabilization on the orbits of small bodies. Near the Sun, sources of such periodicity in the first tens of millions years are not supposed (by [*3; 10; 25*]). That forces us to consider regular inhomogeneities of the Galactic disk in the plane of the Solar orbit as the primary cause of dynamic effects on asteroids orbits [*1; 2*] developed in the form of Spiral Arms [*14*] (*see Conclusion*).

This way, an acceptable causal chain of the relationship between the considered periodically-large-scale processes is established: ME - oxygen and temperatures' drop on the planet's surface - pressure drop - drops in the Ocean's levels and atmosphere towards newly formed impact crater - a large cosmic body impact - destabilization of the orbits of small bodies (asteroids' belts) - passing of shock waves of Spiral arms through the Solar system.

***Other short-time processes which are not directly related to ME.***

***The origin of plumes.*** The diameter of plume channel during rising period remains almost constant, the material passing through the channels in the upper mantle floats mainly vertically and fills the areas located above the channels. In this case, mantle convection does not affect the plume, which begs the question: - How can the fluid which was separated from the liquid core pass through the dense mantle with an allowable speed of 100 cm/year for over 2000 km of the Earth's radius [*6*]? Having dozens of other active zones for "excess" pressure and temperatures unloading in the body of an irreversibly cooling planet? Isn't it the activity of superplumes which caused the mantle "convection"?

There are various opinions about formation of plumes that do not claim universality:

- Due to a violation of the stationary state of the outer core, accompanied by thermal explosions and spontaneous mass emissions of gases. But there is data proves that the content of volatile components in plumes is at the level of gross silicate Earth and even lower.

- Due to the mantle rise of heated jets from the depths, "burning" the lithosphere. But for this, the temperature of the plume must exceed the temperature of the asthenosphere by up to 1500°C [*21*].

- Float-rise into the upper mantle of light material released during the process of density differentiation at the boundary of core and lower mantle or in the D" layer. In this case, the stability of the location of the hot spots on the Earth's surface imposes a significant limitation on substance flowing speed in the lower mantle - their speed should be no more than ~ 1 cm/year, and the range of possible depths of the "roots" of plumes is also limited.

- Penetration, a sudden breakthrough of the subduction substance of the upper mantle into the lower one in the form of global avalanche-like descending plumes ("avalanches"), which forms a compensating global ascending plume. The hypothesis does not explain the mechanism of planetary scale ring structures which is necessary for "avalanche" formation, ignores the debatable "convective subduction immersion" of decompressed upper mantle material in a denser lower mantle, as well as does not apply to plumes with "roots" at depths of up to 100-670 km including plumes existing in superplumes.

- The thermochemical hypothesis suggests that the sources of plumes are mantle funnels at the core-mantle boundary where hydrogen reactions occur. They can form light-fusible components that lower the melting point of the lower mantle substance, which feeds channel and rises the thermochemical plume.

The latest hypotheses admit that the formation of massive ring structures of mantle funnels and supply channels precedes the appearance of plumes, and also serves as a necessary condition and could be a possible reason for their origin. The funnels and supply channels can logically be explained as impact's consequences - in the form of near-surface impact craters of cone-shaped and channel-like forms after high-speed bodies penetration: - In most cases, real craters are different from hemispherical - impact holes at ultrahigh speeds are crater tunnels with a proportional depth to $V^{2/3}$ of the impact projectile, with round-shaped impact craters significantly exceeding the diameters of those impact projectiles (by [22]). On Mars, Callisto and Ganymede, central dimples are observed in the deep parts of impact craters [16], cone-shaped craters and penetration channels of "explosion" burried in the deepest parts, many large plutons and layered intrusions have the same structures. Looking at highly studied kimberlite diatreme we can divide it to the crater part, diatreme part and channel part.

According to the widespread but very limited applicable shock-explosive analogy of quantity, the effective center of the "explosion" caused by impact is located at a depth approximately equal to the diameter of the impactor, but the exact depth remains unclear [16]. The situation becomes clearer if instead of one time "explosion", on fixed depth for all speeds and structures measured as 1-2 diameters, we draw an analogy with the continuous generation of shock wave from the impact point with the target until it stops and the pressure unloads. Comparison with an explosion of a linear explosive charge is acceptable: boreholes (mining) of the same diameter for the most diverse depths and linear explosive charge have funnel cones of emissions of approximately equal depths and diameters. The empty remainder of the borehole channel is deeper than the ejection cone, the so-called "drinking glass" remains without obvious signs of explosion.

According to modern concepts of the shock wave geometry, the following is assumed: - The high-pressure area which was created by ultra-fast impact is a thin shell and is not a hemisphere [16], after the projectile speed exceeds the speed of the shock wave arising in the target's material. The shock wave forming a near-surface crater spreads radially from the axis of the immersion channel of an ultra-high-velocity projectile and damps-dissipates proportionally to the passed distance and increasing lithostatic pressure with depth. Therefore, the near-surface ejection funnel can have less depth than the impactor's penetration depth: - The final shape and size of the craters are practically not related with the characteristics of energy source - the estimation discrepancy of the impactor's energy by the crater's size can be different up to 40 [16].

Questions are caused by the projectile stopping mechanism, proposed by the shock-explosive analogy: - A shock wave is reflected in the form of a rarefaction or unloading wave after it reaches the impactor's surface [16]. Underpressure travels with the sound speed in the compressed material, discharging it to almost zero pressure, and slows it down [16]. During discharging the momentum and energy of the impactor is transferred to the target and ends only when the rarefaction wave reaches the interface [16]. Such an approach can lead to the misconclusion that the higher the sound speed and shock wave in the material of the impactor, the smaller its penetration depth. In fact, after impact the materials of different densities as well as the speed of the shock wave in denser materials are getting higher.

Based on the above-mentioned data, it is advisable to use the extrapolation of experimental data to larger scale events using the principle of similarity. According to the similarity principle for impact velocity of 30 km/s, its internal and kinetic energies of the impactor become almost constant and make up only a small fraction of the total energy, starting from time the impactor reaches a depth of 10 own diameters (by [16]).

To calculate the penetration depth of cannon artillery projectiles we use empirical Berezanskaya and other formulas [23]. These formulas are directly proportional to low-speed (up to 0.6-1.2 km/s) bodies' impacts with different densities by [22]. According to the materials [23], using speeds up to 1.2 km/s the penetration depths of metal ball in a target with a density of up to 2.7 g/cm$^3$ can be greater than 10 projectiles' diameters. With a further increase in the impact velocity, the penetration depth continues to increase, but already in a power-law relationship from $V^n$ (0,3<$n$<0,88) [16; 22] (*Fig.3*).

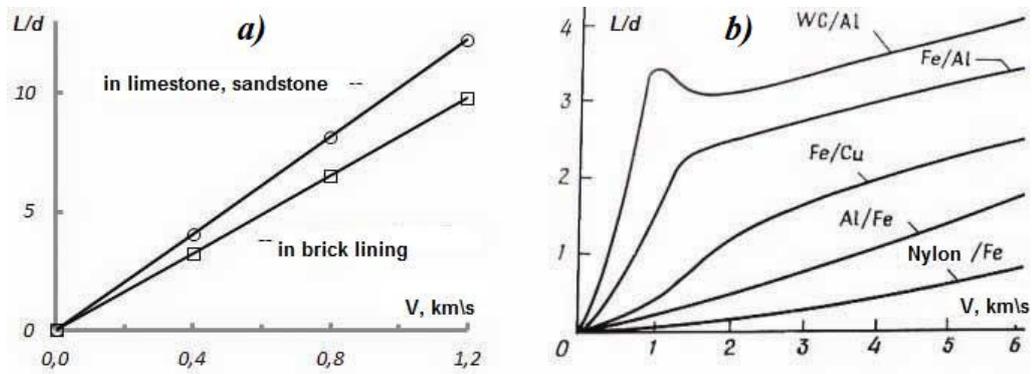

*Fig.3. Penetration depths, related to metal balls' diameter, according to the materials [23] of artillery test results (a), penetration depth, related to impactors' diameters, depending on the impact speed for different combinations of projectile and target materials [22] according to the results of laboratory experiments (b).*

Charters and Summers formula is proposed for calculating the penetration depths in case of collisions of ductile metals at speeds of 1-4 km/s by normal (*A.C. Charters & J.L. Summers, 1959*) $H/L=2,28*(P_p/P_t)^{2/3}*(V_i/c)^{2/3}$, where $H$ - penetration depth of an impactor, $L$ - impactor's diameter, $P_p$ - impactors' material density, $P_t$ - target's material density, $V_i$ - impactor speed, $c$ - sound speed in target's material [*16*]. There are calculation methods for different combinations of materials and impact speeds up to and greater than 14-18 km/s. They all have a similar look $H=K*L*V^n$, where $0,3 \leq n \leq 0,88$, and $K$ - empirical coefficient by (*Fig.3*). The widespread opinion about "explosive" evaporation of impactor's material at a speed of collision of >12 km/s is caused by limits of durabilities of the materials used in experiments: Destruction-softening of projectiles and parts of light-gas guns at moments of acceleration pulses during shots does exceed G-loads at target braking. Projectiles can also be destroyed as a result of pressure unloading, not at the moment of increasing impact load and after passing thin targets at speeds up to 1-6 km/s. And there may be cases where projectiles remain intact after collisions at speed >10 км/s [*22*]. A data confirms that firing a gabbro-anorthosite target at a speed of up to 11 km/s makes the iron projectile melted only partially, and complete metal transition into a vapor state occurs at more than 30 km/s collision speed.

Of course, it includes the difficulties of proportional changes in gravity acceleration and other significant conditions - water intention, viscosity, temperature and increasing lithostatic pressure of the target's rocks which make laboratory experiments applicable only to rare extent comparing with planetary impact processes (by [*16*]). But despite the above mentioned difficulties, the results of experiments for projectiles differing in mass (3-65 g [*22*] and 4.7-113.0 kg according to the materials [*23*]) prove that the gravity acceleration remains constant as well as properties of impacting materials are comparable (*Fig.* 3) which indicates (in diameters of the impactors) increasing penetration depths at speeds up to 1.2 km/s for more than 5 times with an increase by 2-4 times. Such a decrease in ballistic limits (the minimum required densities, sizes, and velocities of projectiles for guaranteed penetration of targets' thickness) confirms the scale phenomenon and suggests that after increasing in the size of experiments up to planetary values, the penetration depth cannot be less than the extrapolated data of small-sized laboratory test. According to the Charters and Summers formula a spherical iron projectile with a diameter of 250 km can normally penetrate the outer core through the mantle with an average radial density of 4.4 g/cm$^3$ and with the sound speed inside the target material (average radial P-wave velocity) is 11.6 km/s, at a speed of 72.8 km/s. The calculated data is fully compatible with the sizes of 250-1000 km for the majority of the channels of cylindrical mantle plumes and velocity inhomogeneities are revealed by modern geophysical methods.

Considering the planet as an obstacle of 12.7 thousand km thick, with an average radial density of 8.5 g/cm$^3$ and a P-wave speed of 10.6 km/s, for the normal collision speed of 72.8 km/s, the maximum size of the impactors can be: iron - 1.6 thousand km, stone - 2.8 thousand km: - The Proportional $V^{2/3}$ dependence is confirmed for projectiles' penetration depths at ultrahigh speeds. The same is for impactors that significantly lower in strength than target material and even for those that are completely collapsed upon impact [*22*] (*Fig.3*). According to the estimated channel sizes of antipodal superplums up to 4.4 thousand km, the ballistic speed limit of an iron asteroid to penetrate a planet-like target with a thickness of only 3 shell diameters can be 16 km/s. The possible existence of a channel through penetration in the body of the planet is evidenced by coincidences for dimensions and extensions of channels of antipodal superplums and their antipodal superexpressions in ground features, different layers of mantle and surface of external core according to the materials [*15*]. As well as the spatial, large-scale and chronological relationship with the largest antipodal ring megastructures - Pre-Atlantic and Pacific edge pits Pantalassa presumably Permian/Triassic (P/T) age.

*Cooling surges.* The main source of heat loss in the planet is ocean, mainly young bark [*10*]. Maximum heat losses compared to mid-ocean ridge (MOR) spreading zones are observed in plum hot spots with a geodynamic situation which is fundamentally different from spreading - high rise rate, increased temperature and productivity of magmatism. Based on the "lag" duration (up to the first tens millions years) of the first existence of the magmatic activity of the plume on the Earth's surface from the moment of its occurrence [*6*], and the rapid (up to 1-3 million years) "formation" of large trap provinces during cooling surges of the planet, it is concluded that the impulses of the heat losses of the planet are not related to the existing plumes, but might be the source of the new ones.

Another hypothetical reason for mass surges in heat losses with growth impulses of the lithosphere is called post-impact isostatic filling of impact craters with mantle material identical to the plume rises (by [*7; 16*]).

*Oxygen generation.* The atmosphere became more than 1/5 oxygenated before the terrestrial plants and this fact forces us to consider the alternative theories for non-organic sources of oxygen in the formation of the Earth's atmosphere: atmospheric oxygen is a product of water decomposition and not carbon dioxide regeneration, in addition, photosynthesis is not able to provide the capacity of all modern oxygen. The sharp increase in oxygen percentage in the Late Paleozoic with a long, possibly up to 15 million years of low biodiversity (by [*1*]), except atmospheric transgression may be due to hard cosmic radiation increase to transgressive/regressive highland intercontinental basins. A long-term massive oxygen and hydrogen increase is possible as a result of thermal dissociation of water in a newly formed depression with a possible initial depth to the surface of the outer core, which currently has a temperature up to 5000°C.

*Turns of the equatorial plane.* Some researchers consider fundamental changes in gravity center of the planet happened after rejecting a large amount of material from the mantle and core, with the following movement of that mass defect to the pole region, and a new gravity center to the equatorial plane (by [*6; 21*]). It should be noted that even the opening of the oceans and the supercontinents accretion (restructuring the locations of the oceans and continents) could not affect the largest moment of inertia of the Earth, depending only on mass and radius [*20*], if we do not allow the instantaneous formation of new crater basins and their greater initial depth. Based on thermodynamic assumptions, the newly-formed and shifted to the polar region massive depression will be gradually filled with deep mantle-nuclear material, forming a positive gravitational inhomogeneity similar to Lunar masks (by [*16*]), with the inevitable displacement of the mass concentration (mascon) anomaly - by rotational force into the equatorial region [*20*].

Obviously, for a significant turn of rotation axis of a huge gyroscope, with a significant difference in polar and equatorial radii, a cardinal changes in the mass distribution in the newly generated form of the Pacific Basin-supermascon may be required. But taking into account the possibility of independent rotation of the mantle layers and the core, as well as relative to each other along the boundaries of phase transitions where the material plasticity can sharply increase, for the considered motions it is possible to expect that the changes in the upper mantle are sufficient.

*Outbreaks of tectogenesis.* The classical theory of mantle convection is unacceptable for explaining the movement of oceanic platforms [*10*], therefore, in conjunction with data on the spreading's lull at geochronological boundaries [*12*], "convection" cannot be considered as a reason for the activation and the continental tectogenesis or shifting the planet's layers. Many researches consider the rotation factor that changes the axis and speed of rotation of the planet or its individual layers as the main cause for impulses of tectonic activity [*10; 13; 20*]. Since the largest principal moment of the Earth's inertia depends on mass, radius [*20*] and mass distribution, the causes of the necessary changes can be considered as massive impact emissions of mantle material with a decrease in mass and a change in the gravity center of the target as a whole mantle cover, as well as post-impact size reductions of the planet during magmatic filling the crater, with a repeated change in the center of mass. The superimposed acceleration of the Earth's rotation, with a continuing decrease in size, has been noted by individual researchers nowadays.

*Inversions.* The short-term phases of the geomagnetic inversions increasal include rapid 180° pole migrations almost in a circle, relatively close to the 90° meridian in the Western and the Eastern hemispheres. This circle is subper-pendicular to the axis of antipodal rising African and Pacific superplumes, coincident with the equatorial swellings of anomalous masses joints in the planet's cover - the stable axis of the main moment of inertia of modern Earth (by [*15; 19; 20; 21*]). Therefore, unlike the turn of the equatorial plane with a change in the axis of rotation and the dipole of the main moment of inertia, the case of inversion considering the mass dipole retains its rotation in the ecliptic plane but the axis of shell rotation and the equator return to their previous relatively to the Sun orientation. In this case, a pole changing is possible as a result of smaller, gradually accumulating changes in the mass balance of the upper mantle shell that is unstable during the redistribution of mantle material as well as by other large impacts.

*Scheme of possible cause-effect relationships of boundary events.*

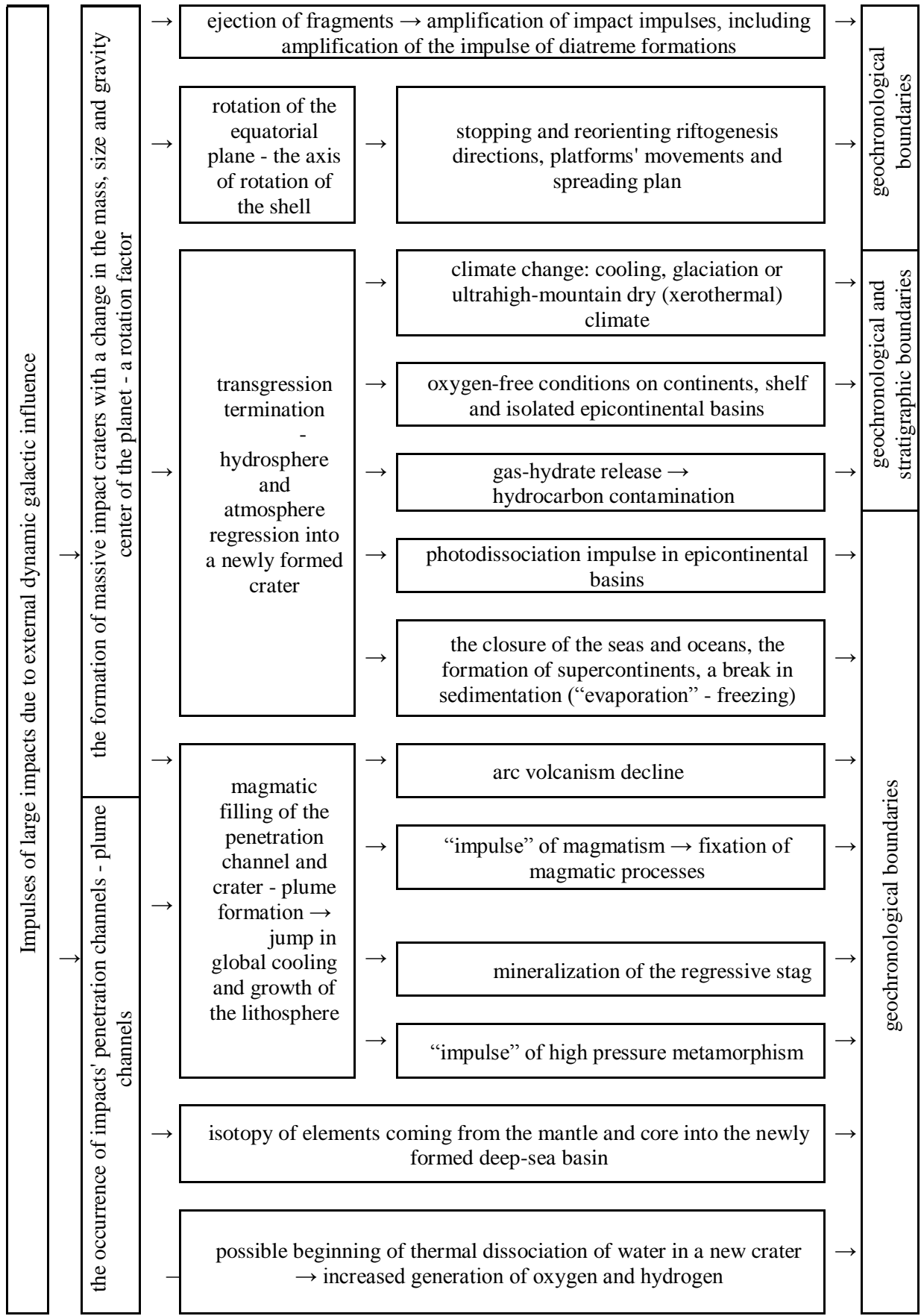

Thus, almost all chronologically and massively related global fast-moving processes on the Earth can be reasonably combined into a single consistent system where the impulses for large asteroids' fall originated by external galactic effects in the flow of gradual long-term processes. It confirms Halley's hypothesis (*E. Halley, 1694*) of impact nature for deep-sea basins formation and global disasters in the history of the Earth. The complex of interrelations is consistent as both with established regular periodicities and chronological sequences of sharp changes that coincidences with the intensities of boundary events and the scale of subsequently gradual processes. The existence of a system can be a sufficient evidence for unified nature of short-term processes (by [*1; 10; 13*]), as well as impact hypotheses of the formation of ring deep-sea basins and plumes, multi-kilometer regressions and the rotational source of earthquakes, as the only possible conditions for the existence of the system.

**Conclusion.**

*Galactic origin.* The existence of a periodicity for 30 to 12 million years in the Phanerozoic history of presumably external dynamic influences on the objects of the Solar system is possible only with rotation of established regular disturbances-inhomogeneities of the Milky Way disc - the Spiral Arms, with a significantly exceeding angular velocity from the Sun's angular velocity. The estimated turnaround time for a model of two main Arm-disturbances (as the duration of two epochs of the geochronological scale) lasts for about 26 million years in the last 90-100 million years (according to [*1*]) and 23.5-23.8 million years at present (*Table 1; Fig.2*), which fits the data on the current circulation period of the Galactic center objects at about 24 million years.

The possibility of existence of a stationary galactic shock wave (GSW) does not mean that it must exist. Lagging density waves, emissions from the Center of gases, explosive products or other matter (energy) can be responsible for the Spiral structure of the Galaxy (by [*14*]) which velocity is spread from the center to the periphery. In this case, the angular velocity for disturbances spread on the environment will correspond with the rotation speed of the dipole radiation generator itself (traditionally this mechanism is compared with rotating of irrigation system, less with Parker spirals of the Solar wind), significantly exceeding the angular rotation velocities for objects of the Galactic plane which are outside the central "corotation" zone with a radius of about 3-4 kpc (*Fig.4*). The established differences in the characteristics of spiral disc inhomogeneities [*14*] will depend on the properties (velocities) of possible outliers. According to the Galaxy model [*4*] (*Fig.4*) and the current rotation periodicity of the objects of the Galactic Center at 23.5-24.0 million years, the radial components of the Spiral disturbances extensions can be approximately estimated as following: Sagittarius and the Norma-Outer Arm at 350-520 km/s, Perseus and the Scutum-Centaurus at 450-550 km/s, Orion (Local) and its Antipode (*by T.M. Dame, P. Thaddeus, 2011*) at 600-800 km/s. The indicated speeds correspond to the stellar wind range of E.N. Parker and ranges about 100-1000 km/s and the observed extremes of solar cosmic rays (SCR) at 260-500 km/s, 500-1000 km/s and up to 800-1500 km/s.

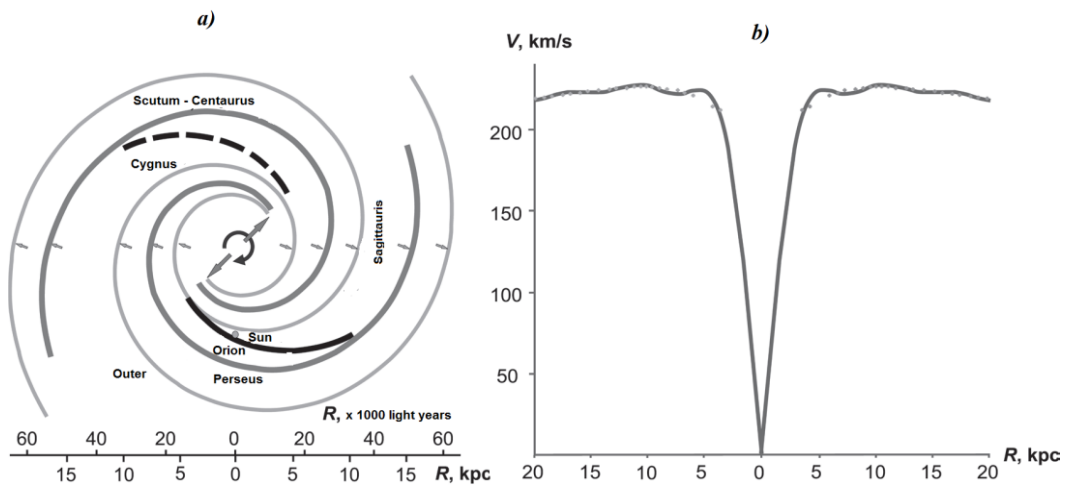

*Fig.4. Diagram of the logarithmic spiral arms of the Milky Way according to the data [4] (a), galaxy disk rotation curve [5] (b).*

Streams of star rays to $E > 10^9$-$10^{10} eV$ (ionized plasma mainly from protons) correspond to the characteristics of the expected emissions, they have a shock wave and the necessary energy. With a high radiation intensity (the stellar density in the central part of our Galaxy is 4-7 times higher than in the

"galactic" surroundings of the Sun). The possibility of particle accelerations at GSW fronts and in the magnetic field of the core, the long-term pressure of high-speed stellar CRs of high energies from the Center seems sufficient to destabilize the orbits of small bodies of the Solar System, as well as for stopping the flow. The real spiral structure represents both a density disturbance and a velocity disturbance in the galactic disc [14] which is up to 10-15% (*Fig. 4b*).

SCR shockwaves, observed in near-Earth space in most cases associated with non-stationary sporadic high-speed flows. There is an evidence that the energy density of high-speed solar wind can increase tens, hundreds and thousands of times compared with calm state and reach values of 100 n/cm$^3$ and 50 Erg/cm$^2$·c. By analogy with sporadic SCR, the massive boundaries of geological eras may well correspond to the passage through the Solar system of high-speed fluxes of the Orion (Local) Arm and its Antipode, which is consistent with the modern CR anisotropy of ultrahigh E>10$^{18}$eV from the center of the Galaxy. Lower-speed arms - the boundaries of centuries are represented less in the geochronological scale's history.

According to the existing ideas after ionization the stellar wind can be ejected only along the magnetic axes of supermassive objects, similarly to ejections of narrowly directed radiation beams from the magnetic axes of pulsars that do not coincide with the axes of rotation. The arms of most spiral galaxies which lie strictly in the plane of rotation may indicate the coincidence of the magnetic axes of their nuclei with the planes of rotation. Apparently, the interacting magnetic fields of close binary systems of supermassive objects ("black holes") merge into a single powerful magnetic dipole, coinciding with the mass dipole long before the actual merging of the bodies themselves. At least two objects with masses up to 10$^6$ Solar masses are supposed to exist in the core of the Milky Way (by [*14*]), falling with acceleration at each other. Logarithmic kind of the Sleeves which are lagging behind the rotation of the core (by (*H.W. Babcock, 1938*) unlike the Archimedean-Parker's spiral. In the case of steady rotation of the emitter, confirms the acceleration of rotation of the Galaxy core as a merging binary system and is consistent with an increase in the frequency of boundaries - a decrease in the duration of the Phanerozoic eras, presumably by 2.8 million years every two geological epochs, or per turnover of the Galaxy core by (*Fig.2; Chart.1*).

If the Orion (Local) Arm width near the Sun is about 3 thousand light years (by [*4*]), the possible time for the galactic cosmic rays to pass through disc objects with a radial flow velocity (calculated from [*4*] 600-800 km/s) could be about 1.1-1.8 million years. During this time, stellar rays of permissible density can transfer a considerable amount of energy to disc stars, their volume depends only on the dimensions of the receivers. According to available hypotheses, giant stars, owing to their sizes, can accumulate in the outer layers the maximum of additional energy in a relatively short time which probably could make up to critical 10$^{49}$-10$^{51}$ erg (10$^{38}$-10$^{44}$ Joules) that are needed for SNII type supernovae outbreaks. Supernovae of this type are typical only for Spiral Arms, which confirms the admissibility of the stellar-wind nature of above-mentioned radiations. Collisions-fusions of the stars are called one of the possible causes for flares of another type.

Hereby, the assumption about the root cause of historical global-scale impulses of changes on the Earth does not contradict the modern data concerning acceleration of the Galactic core rotation and confirms the existing valid hypotheses about the Spiral branches' extensions resulted after bombardment the asteroid belt by large bodies as part of periodic dynamic disturbances in the medium of the Milky Way disc followed by expanding from the center streams of stellar winds along the magnetic axis. Using data about the inconsistency of the Milky Way's curve rotation with the laws of mechanics, observations of radial movements of stars, dust and gas flows from the Galactic Center allows us to consider the described mechanism as a possible reason for existence of the disc itself.

*The author is sincerely grateful for the information support, methodological recommendations and constructive criticisms to A.M. Anufriev, A.H. Abdulov, A.V. Bagrov, A.T. Bazilevsky, A.A. Barenbaum, S.V. Belov, S.V. Bykov, I.S. Veselovsky, N.M. Gladkykh, G.A. Goncharov, A.V. Ermokhin, A.V. Zakharov, B.S. Zeilik, I.I. Zinchenko, B.A. Ivanov, A.N. Kozlovsky, E.M. Korobova, V.V. Kuznetsov, G.N. Kuzovkov, Yu.A. Lavrushin, A.V. Lobanov, O.Yu. Malkov, V.L. Masaitis, Yu.P. Orovetsky, L.V. Rykhlova, S.I. Svertilov, V.I. Sergienko, A.Ya. Sidorin, G.Ya. Smolkov, V.I. Starostenko, V.G. Surdin, S.D. Trankovsky, A.N. Tyurin, V.I. Feldman, V.M. Fomin, A.I. Khlystov, A.M. Cherepaschyuk, R.F. Cherkasov, V.V. Shevchenko, A.A. Shylo, N.I. Shishkin, G.A. Shishkina .*

*Published in the journal "Otechestvennaya geologiya" [National geology], 2015, №3. pp.70-83.*
*Translated by R.M. Garipov*